\documentclass[11pt,twoside]{article}
\usepackage{asp2010}
\usepackage{epstopdf}
\resetcounters

\begin{document}

\title{aTmcam: A Simple Atmospheric Transmission Monitoring Camera For Sub 1\% Photometric Precision}
\author{Ting Li,$^1$  D. L. DePoy,$^1$ R. Kessler,$^2$ D. L. Burke,$^3$ J. L. Marshall,$^1$ J. Wise,$^1$
J.-P. Rheault,$^1$ D. W. Carona,$^1$ S. Boada,$^1$ T. Prochaska,$^1$ R. Allen$^1$
\affil{$^1$Department of Physics and Astronomy, Texas A \& M University, 4242 TAMU,College Station, TX 77843, USA}
\affil{$^2$Department of Astronomy \& Astrophysics, University of Chicago, 5640 S. Ellis Ave,Chicago, IL 60637, USA}
\affil{$^3$SLAC National Accelerator Laboratory, Menlo Park, CA 94025, USA3}}

\begin{abstract}
Traditional color and airmass corrections can typically achieve $\sim$0.02 mag precision in photometric observing conditions. A major limiting factor is the variability in atmospheric throughput, which changes on timescales of less than a night. We present preliminary results for a system to monitor the throughput of the atmosphere, which should enable photometric precision when coupled to more traditional techniques of less than 1\% in photometric conditions. The system, aTmCam, consists of a set of imagers each with a narrow-band filter that monitors the brightness of suitable standard stars. Each narrowband filter is selected to monitor a different wavelength region of the atmospheric transmission, including regions dominated by the precipitable water absorption and aerosol scattering. We have built a prototype system to test the notion that an atmospheric model derived from a few color indices measurements can be an accurate representation of the true atmospheric transmission. We have measured the atmospheric transmission with both narrowband photometric measurements and spectroscopic measurements; we show that the narrowband imaging approach can predict the changes in the throughput of the atmosphere to better than $\sim$10\% across a broad wavelength range, so as to achieve photometric precision less than 0.01 mag.
\end{abstract}

\section{Introduction}
The discovery of the accelerating universe ranks as one of the most important discoveries in Cosmology in the past decades. Within the framework of the standard
cosmological model, this implies that about 70\% of the universe is composed of a new,
mysterious dark energy; there is yet no persuasive theoretical explanation for its existence or magnitude. The Dark Energy Survey (DES) is aimed at improving our understanding of this mystery  \citep{2006astro.ph..9591A} using multi-band imaging over $\sim$5,000 square degrees with DECam, an extremely red sensitive 520 Megapixel camera. DECam will be installed at the prime focus of the Blanco 4-meter telescope at Cerro Tololo Inter-American Observatory (CTIO), a southern hemisphere NOAO telescope.The DES has a goal of reaching 0.01mag photometric precision in DES-grizY band to achieve the science goals of the survey.\footnote{More details at http://www.darkenergysurvey.org/} Recent studies have shown that precision determination of various dark energy parameters can be improved with better photometric precision. The systematic uncertainty of the measurements of the dark energy equation of state parameter, $w$, from the first three years of the Supernova Legacy Survey (SNLS3), for example, is dominated by the photometric precision of the survey \citep{2011ApJS..192....1C}. In addition to Cosmology, other forefront science issues also demand better performance on photometry.

It is difficult to achieve ground-based photometry measurements with precision below 1\% for large surveys over a kilo-degree$^2$ area. The Sloan Digital Sky Survey (SDSS), for example, achieved relative photometric calibration reproducible to only 1-2\% (rms) over the entire survey field. Unmodeled atmospheric variations are responsible for almost all of the calibration error budget for SDSS \citep{2008ApJ...674.1217P}. Variations in the wavelength dependence of atmospheric transmissivity can induce systematic errors that depend on source colors. \citet{2007AJ....134..973I} found that, assuming a standard atmosphere, using synthetic photometry for stars from the Gunn-Stryker atlas, this effect can induce offsets of up to $\sim$0.01 mag for the $u-g$ and $g-r$ colors when air mass is varied by 0.3 from its fiducial value of 1.3. Moreover, potentially larger errors could be induced even at a constant air mass if the wavelength dependence of atmospheric transmissivity is significantly different from the assumed standard atmosphere; this will be discussed substantial in the next section.

These studies suggest that unmonitored changes in the atmospheric throughput ultimately limit survey photometric precision to $\sim$0.01-0.02 mag. Therefore, we propose to build an Atmospheric Transmission Monitoring Camera, called aTmCam, to monitor the atmospheric transmission in real-time, as an auxiliary system for the Dark Energy Survey. The system would be installed at Cerro-Tololo Inter-American Observatory(CTIO), and could in principle be used to calibrate data from all instruments at the observatory.

This paper is organized as follows: In section 2, we discuss how the atmospheric variation induces photometric errors and then determine the requirements for the atmospheric throughput measurement. In section 3, we describe the method to measure the atmospheric throughput and the proposed instrumentation system; we also introduce the prototype system and present the preliminary results. In section 4, we show the plan for the next test at CTIO this coming fall, and then we conclude.

\section{Requirements}

Atmospheric transmission in the DES wavelength range ($\sim$300nm-1100nm) is mainly determined by three radiative processes in the atmosphere \citep{2007PASP..119.1163S}: Rayleigh scattering from the molecules, which is simply dependent on the barometric pressure; aerosol and dust scattering from small particles and; molecular absorption, in particular by O$_2$, O$_3$, and H$_2$O. The O$_2$ lines are saturated and the so-called ``rate of growth" is closely proportional to the square root of the barometric pressure, so it can be computed and scaled with the Rayleigh scattering appropriately.

\begin{figure}[h]
\plotone[width=0.6\textwidth]{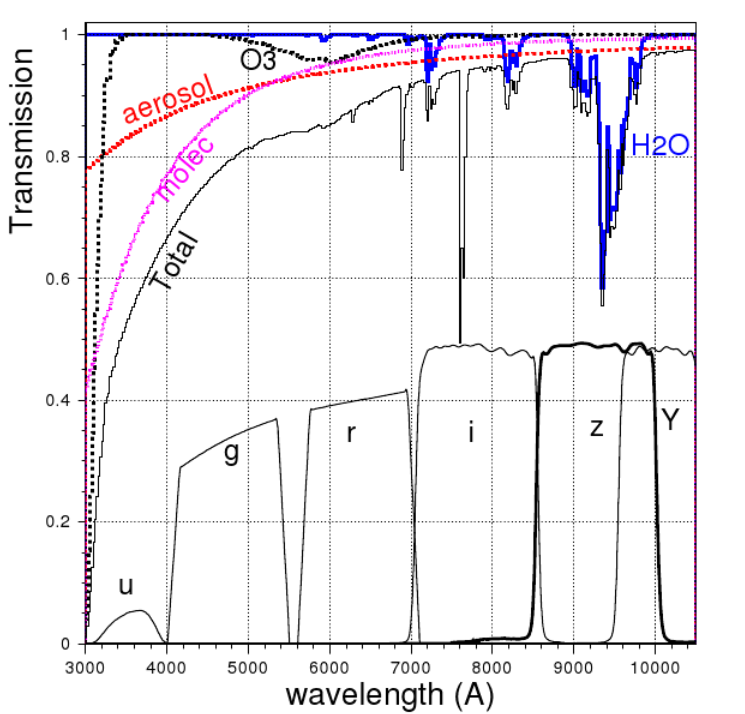}
\caption{Typical normalized atmospheric throughput above CTIO showing the major sources of attenuation by the atmosphere, generated by Modtran. The normalized pre-construction estimates of the DES filters are also shown except that u-band is actually a SDSS-u' filter since DES-u is not yet defined.
\label{fig:1}
}
\end{figure}

We investigated the impact of atmospheric variations on the DES calibration with the variations in the four main components: precipitable water, ozone, molecular and aerosol scattering. Figure \ref{fig:1} shows the transmissivity of each component from a fiducial atmosphere over CTIO and the various DES bands. The DES-grizY bands conform to the anticipated performance of DECam on the Blanco telescope. DES has not yet decided to add a DES-u filter, so SDSS-u' filter is shown as a potential ultraviolet filter. For the remainder of the paper these bandpasses will be referred to as ugrizY, although they are slightly different from other photometric systems.

\begin{figure}[h]
\plottwo{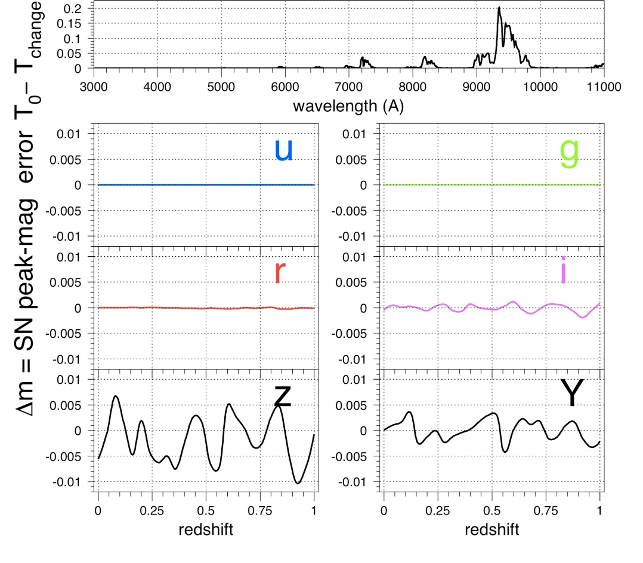}{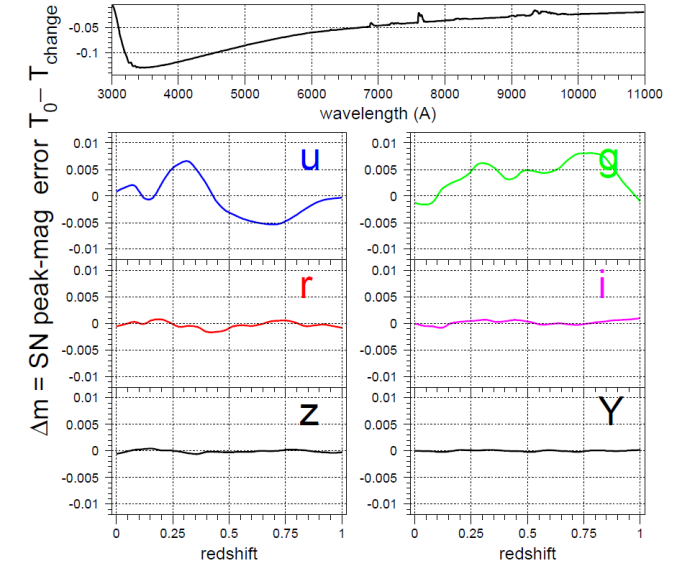}
\caption{Left panel: Top figure shows the fractional difference between the original atmospheric transmission and one with additional precipitable water.  Bottom figures show the systematic error in SN photometry introduced by this change in the water absorption in different bandpasses. Right panel: Similar to the left panel but with lower aerosol optical depth. Note that these errors are residuals after standard calibration techniques are applied.
\label{fig:2}
}
\end{figure}
Our simulation calculated the synthetic ugrizY photometry for an object-SED using a model of atmospheric transmission corresponding to a photometric night. The same object is then simulated using an atmospheric transmission in which one of the components has changed significantly. We added more precipitable water so that the atmospheric transmission caused by water absorption was 20\% deeper around 940nm (see the top figure of the left panel in Figure \ref{fig:2}).  We then calculated the errors induced by this change in the photometric precision of Type Ia Supernovae at various redshifts (left panel of Figure \ref{fig:2}). Note that this result assumes an initial successful calibration based on the use of standard stars to determine photometric zeropoints and the nightly color and airmass corrections. The systematic error in z-band photometry is approximately ±0.01 mag, which is the entire DES photometric precision goal. The primary reason is the variation in expected H$_2$O absorption around 940nm.  There are significant effects on i and Y band as well. If we could monitor the water absorption better than $\sim$10\% (that is, measure the depth of the water absorption around 940nm with an accuracy of $\sim$10\%), we then could achieve photometric precision less than 0.005 mag.

The H$_2$O water absorption at CTIO varies significantly over the expected range of conditions. In particular, the amount of precipitable water varies from $\sim$0mm to 9mm and the average is around 3-4mm over the year \citep{1975PASP...87..935H} . We note that \citet{1975PASP...87..935H} see factors of 2-4 changes in the precipitable water on timescales of $\sim$days during their 3-year study.

Similar errors are introduced into u- and g- band photometry by a significant variation in aerosol scattering in the simulation.  For example, if we decrease the optical depth of the aerosols in the atmosphere so as to cause a $\sim$15\% change in the atmospheric transmission around 350nm (top right panel of Figure \ref{fig:2}), photometric precision errors will be approximately ±0.005 mag in u-band and +0.008/-0.002 mag in g-band. Again, these account for large fraction of the DES photometric precision goal. If we could monitor the aerosol scattering better than $\sim$10\% (that is, measure the depth of the aerosol scattering around 350nm with an accuracy of $\sim$10\%), we then could achieve photometric precision less than 0.005 mag.

O$_3$ can cause small changes in the atmospheric throughput, particularly around 600nm and below $\sim$330nm. However, over the expected variation in the amount of atmospheric ozone, the difference is small and will be well determined by standard calibration techniques. Unless DES creates a "DES-u" filter that includes significant transmission below $\sim$340nm, the ultraviolet absorption of ozone should have an insignificant impact on photometric precision. There is also significant variation in the Rayleigh scattering component of the atmospheric throughput over the expected range of pressures at CTIO. However, the strength and profile of this component is predicted by the local atmospheric pressure, which is easily monitored.

To summarize, a system to monitor the atmospheric throughput above CTIO will need to measure the changes in the throughput of the atmosphere to better than $\sim$10\% across a broad wavelength range (mainly the changes from the precipitable water and the aerosol optical depth), to achieve photometric precision less than 0.005 mag. Other quantities that will have to be monitored include atmospheric pressure (which will allow determination of the Rayleigh scattering component) and zenith angle (airmass) of the target.

\section{Instrumentation System}

We have turned the requirements given in the previous section into a system design. In this section, we describe the system, and present the setup of the prototype system and preliminary results from the prototype.

\subsection{Philosophy}
The spectrum of an astronomical object observed from Earth is the spectral energy distribution (SED) of the object convolved with both the atmospheric throughput and the instrument response function. We therefore can derive the atmospheric transmission with suitable observations of calibration stars using well-calibrated instrumentation. The calibration stars are expected to have the SED known already, such as spectrophotometric standards, white dwarfs, or well-modeled main sequence stars.

Spectroscopic observations of relatively bright standard stars at $\sim$5 minute cadence, over a range of airmasses, and at wavelengths of 400-1000nm can produce high quality atmospheric absorption profiles \citep{2010ApJ...720..811B}.  While this approach is ideal, it has a major drawback in that it requires a high level of personnel commitment to align a telescope with a relatively small aperture ($\sim$10 arcsec) on the target stars, a relatively large telescope, and a stable spectrograph.

We advocate a simpler system, the Atmospheric Transmission Monitoring Camera (aTmCam). This system uses a set of imagers with different narrow-band filters that monitor the brightness of suitable calibration stars. The imagers will have a field-of-view and aperture large enough to enable automatic pointing and tracking of a catalog of stars (i.e. robotic operation). Each narrowband filter will be selected to monitor a different aspect of the atmospheric transmission, e.g. precipitable water, ozone, aerosol and molecular scattering, as described above. With the prototype described below, we show that imaging through several narrow-band filters will be sufficient to constrain an atmospheric model that will be precise to better than 10\% across the DES filter range, which will yield better than 0.01mag photometric precision. We assume that our goal will be to determine the relative transmission across the DES wavelength range; observations with DECam (camera for DES) itself will be used to determine exposure-by-exposure grey terms, for example, and standard photometric calibration procedures for multi-color extinction terms and color-corrections will be in place.

After we take the simultaneous multi-band images of calibration stars with aTmCam, we will select the best fit atmospheric model constrained by the measured color indices (or flux ratios) from a database with a discrete set of models. A more detailed description for identifying a suitable atmospheric throughput model is as follows:

1)	First, we produce a database, generated by an atmospheric throughput program (such as libRadTran\footnote{libRadTran is freely available at http://www.libradtran.org/doku.php}), containing the possible atmospheric throughput models for CTIO varying all parameters in reasonable steps. The parameters include the precipitable water, aerosol optical depth, barometric pressure and ozone (optional). The range and the grid space of the parameters will be determined by the further test at CTIO in the future. From each atmospheric transmission model, the synthetic color indices will be calculated.

2)	Second, we calibrate the aTmCam instrument system (including filters, optics, and detectors) for each of the several narrowband imagers. DECal \citep{2010SPIE.7735E.201R}, the spectrophotometric calibration system for DECam, is good example of the calibration system that could produce suitable calibration measurements. This step needs to be repeated periodically to determine if there have been changes to the system throughput.

3)	aTmCam will measure a selected star simultaneously through the narrow bandpass filters. We then determine the color indices of the atmospheric transmission from the observed images after removing the instrument throughput and the SED of the selected star. (Below we will use Vega (A0V) as our standard star since the absolute flux of Vega is reasonably well measured. However, we can observe any star if we know the spectrum well and it has adequate flux from UV to NIR.) Then we sort through the database and find the atmospheric transmission model that matches the color indices of the measurement. We currently plan to execute the measurements every $\sim$5-10 minutes all night at a range of positions on the sky. This will allow determination of the atmospheric throughput on a relatively fast timescale. A set of stars should be selected that allow reasonable sky coverage, so that the spatial variations in the throughput can be estimated throughout the night.

\subsection{Prototype System}
We have deployed a prototype system to test the notion that an atmospheric model derived from a few color index measurements of stars with known spectral energy distributions can be an accurate representation of the true atmospheric transmission. The system mimics the characteristics of the complete setup mentioned above, but on a smaller scale. The prototype was used to measure only bright stars (principally Vega and Sirius) and uses a smaller aperture (effectively only 40mm). We coupled the measurements of the stellar narrowband photometry with simultaneous observations of the spectrum of the same star with a spectrograph. We derived the best-fit atmospheric throughput model from these two independent measurements. The two were found to agree to better than $\sim$10\%.

\subsubsection{Setup}

\begin{figure}[h]
\plotone[width=0.8\textwidth]{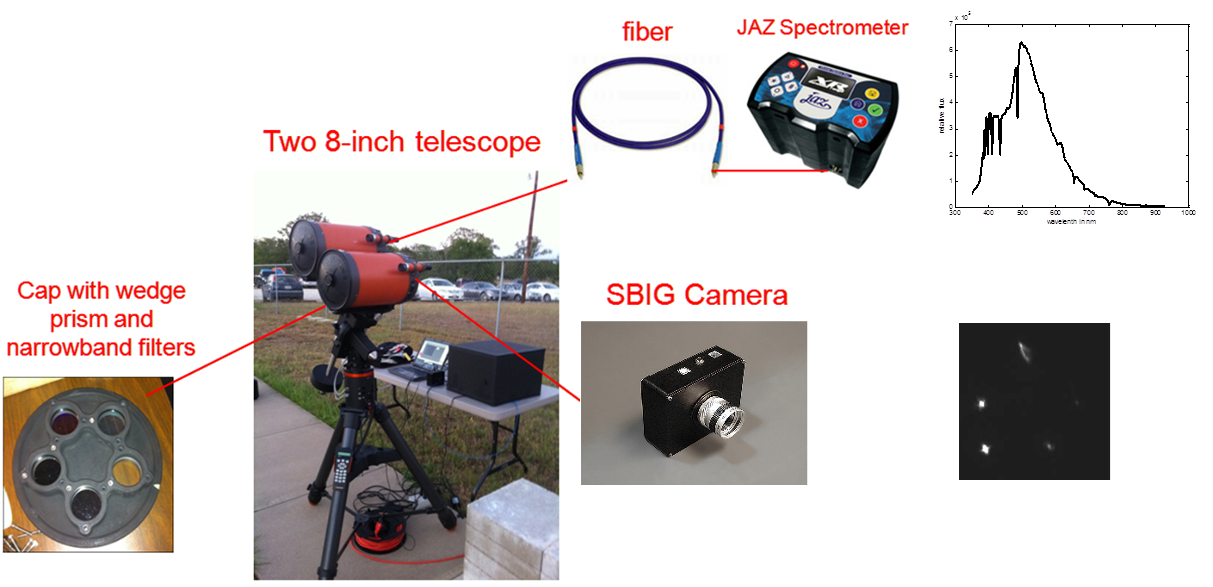}
\caption{Setup of the aTmCam prototype
\label{fig:6}
}
\end{figure}

Figure \ref{fig:6} shows the prototype setup. There are two 8-inch telescopes: one with a fiber at the focal plane that feeds an Ocean Optics JAZ spectrograph, another with an SBIG ST-402ME CCD at the focal plane. The imaging telescope has a cap on the front that holds five filters, each coupled to a "wedge prism" that diverts a $\sim$40mm part of the pupil $\sim$2 arcminutes. This creates five individual images on the CCD of the same star. The two telescopes are co-mounted on a tripod and are aligned to look at the same star, so we simultaneously obtain a spectrum and five narrow-band images of a star. The raw samples from the camera and the spectrograph are shown in the right column of Figure \ref{fig:6}. The relatively poor image quality had a significant impact on the photometric results (particularly the accuracy of the UV measurement). Nonetheless, we were able to generally confirm the performance of the system (on bright stars) and better define the requirements on the final imaging system configurations.

\subsubsection{Filters}

\begin{figure}[h]
\plotone[width=0.6\textwidth]{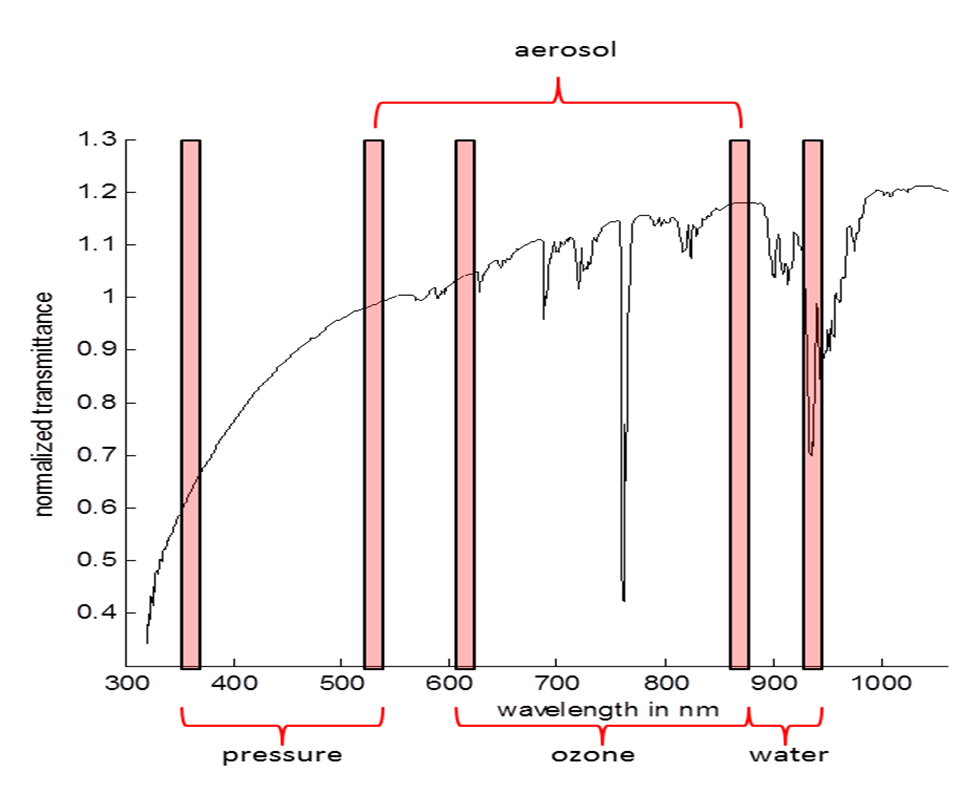}
\caption{Wavelengths of prototype filters relative to a representative atmospheric throughput model.
\label{fig:4}
}
\end{figure}

We initially selected five filters that span 390 $<\lambda<$ 940nm for the prototype monitoring system, since from the five filters we can define four flux ratios or color indices: flux in one filter divided by that in another, and there are mainly four components in the atmosphere that vary.  These filters were not optimized to match the atmospheric transmission features, but were simply close to what seemed like good choices to monitor the precipitable water, aerosol, ozone and Rayleigh scattering components of the atmosphere. They were also available from Edmund Optics and Astrodon at relatively low cost. These filters had central wavelengths and bandpasses as shown in Table \ref{tab:1}. Figure \ref{fig:4} also shows the central wavelengths of the filters on a fiducial atmospheric throughput model, with the indications of the flux ratio from two bandpasses for each component.
\begin{table}[h]
\caption{Central wavelengths and bandpasses of the prototype filters}
\centering
\begin{tabular}{>{\centering}m{4cm}|>{\centering}p{3cm}|>{\centering}m{4cm}}
    \hline
    \hline
    Filter Central Wavelength & Filter FWHM & Parts \# (from Edmund Optics if not specified)
\tabularnewline
\hline
    390 nm & 50 nm & u'2-50R (from Astrodon; equivalent to SDSS-u')
\tabularnewline
    520 nm & 10 nm & NT65-215
\tabularnewline
    610 nm & 10 nm & NT65-225
\tabularnewline
    852 nm & 10 nm & NT65-242
\tabularnewline
    940 nm & 10 nm & NT65-246
\tabularnewline
\hline
\hline
\end{tabular}
\label{tab:1}
\end{table}

\begin{figure}[h]
\plottwo{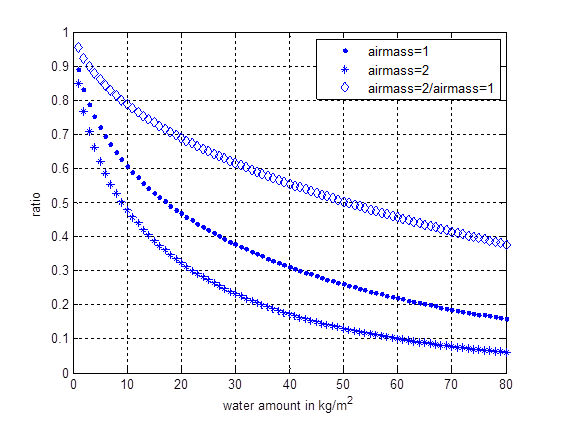}{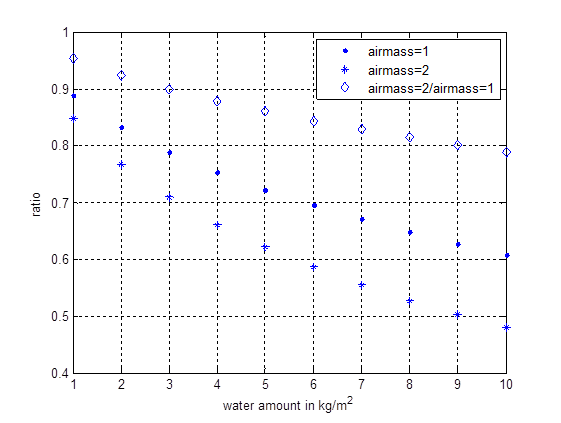}
\caption{Left panel: plot of ratio (defined in the text) as a function of precipitable water; other conditions held constant. Right panel: zoom of left panel. Assuming the amount of the precipitable water at CTIO varies from 1mm to 10mm, the ratio (defined in the text) changed from 0.88 to 0.61 at airmass=1, which corresponds to a $\sim$30\% change compared to the average.
\label{fig:5}
}
\end{figure}

We take water absorption as an example here to show the simulated performance of the selected filters. The ratio from the flux in the 940nm filter relative to that in the 852nm filter is sensitive to the water absorption. Figure \ref{fig:5} shows the results of modeling the expected flux ratios over the range of conditions experienced at CTIO. The ratio changes by $\sim$30\% (depends on the airmass of the measurement) over the expected range of anticipated precipitable water amount at CTIO.

\subsubsection{Instrument Throughput Calibration}

\begin{figure}[h]
\plottwo{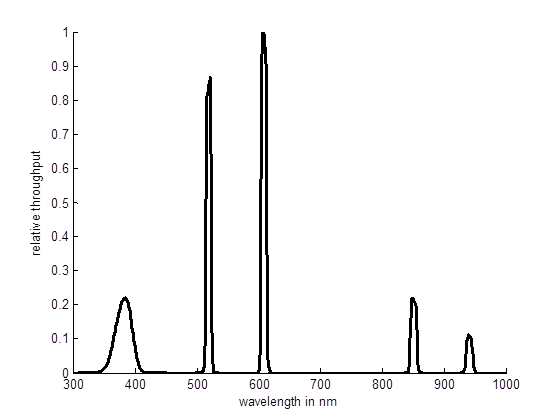}{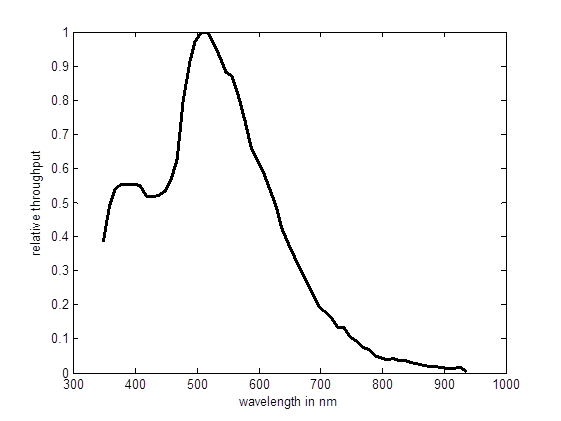}
\caption{Left panel: Relative throughput of the imaging system (cap +  telescope + SBIG CCD). Right panel: Relative throughput of Jaz Spectrometer and telescope.
\label{fig:7}
}
\end{figure}

The throughput of both the imaging system through each filter and the spectroscopic system was measured with a prototype DECal-like system. The throughput was found to be relatively low in the far-red, particularly beyond $\sim$800nm; this is largely due to the response of the CCD. Calibration results for two systems are shown in Figure \ref{fig:7}.

\subsubsection{Preliminary Results}
We have made multiple photometric and spectroscopic measurements simultaneously with the prototype system described above at both Texas A\&M Physics Observatory\footnote{Department of Physics \& Astronomy Teaching Observatory, Texas A\&M University , College Station, TX. http://observatory.tamu.edu/} and McDonald observatory. We use one measurement from each observatory as an example hereafter. We used Vega (A0V) as our standard star since the absolute flux of Vega is reasonably well determined. However, we can observe any star if we know the spectrum and it has adequate UV and NIR flux.

\begin{figure}[h]
\plottwo{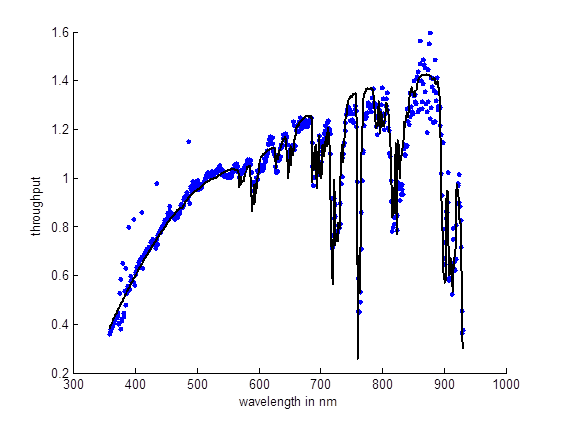}{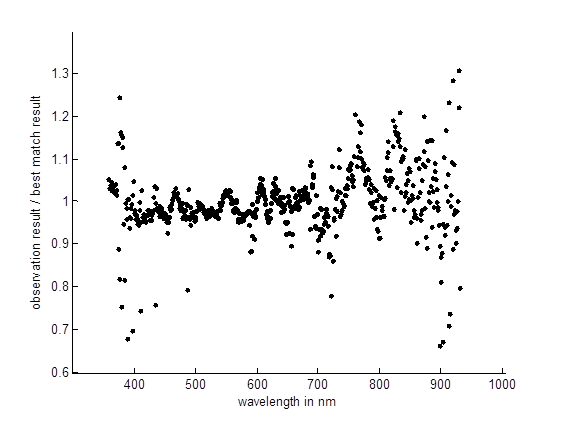}
\caption{Left panel: the measured atmospheric throughput (blue dots) from the observed spectrum and best fit model (black lines) from the synthetic color indices of that measured atmospheric throughput. Right panel: ratio of measured atmospheric throughput and selected model from the synthetic color.
\label{fig:8}
}
\end{figure}

\begin{figure}[h]
\plottwo{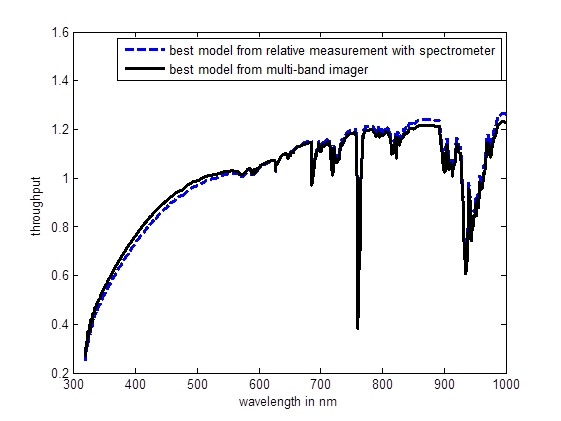}{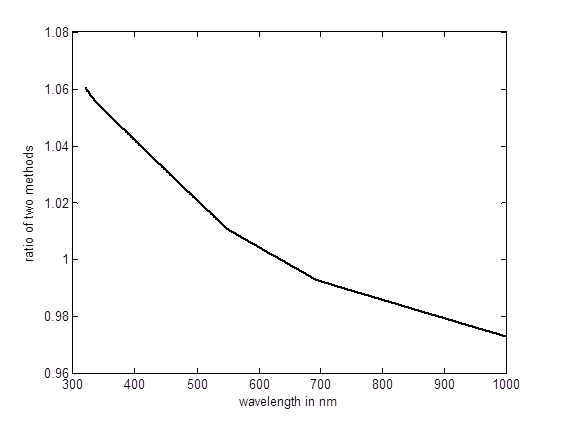}
\caption{Left panel: best fit model from the spectroscopic measurements (dashed blue line) and that  from narrow-band imaging(black line). Right panel: ratio of the two models.
\label{fig:9}
}
\end{figure}

We obtained a set of test observations of Vega on July 12, 2011 at Texas A\&M observatory on Vega. We derive the atmospheric throughout from the observed spectrum taken by the Jaz spectrograph, by removing the instrumental throughput and the Vega SED; this is shown as the dots in the left panel of Figure \ref{fig:8}. We used the SED of Vega from the spectrophotometric standards found in the STScI CALSPEC database\footnote{available at  http://www.stsci.edu/hst/observatory/cdbs/calspec.html}. Images from the narrow-band imaging system had relative poor Signal-to-Noise Ratio (S/N), particularly with the 940nm filter because of the high amount of the precipitable water at Texas A\&M observatory. This is, of course, consistent with the ridiculous humidity we often experience in eastern Texas. We, therefore, did not use the color indices measured by the narrowband imager. Instead, we used the synthetic color indices from the atmospheric throughput measured by the spectrograph (i.e. synthetic flux from those dots in the left panel of Figure \ref{fig:8}), but we synthesize the 910nm bandpass instead of the original 940nm for higher S/N. Then we found the atmospheric model from the database matched best with the derived color indices (using a chi-square fit), which is shown with black line in left panel of Figure \ref{fig:8}. The selected model has residuals compared to the actual measured atmospheric throughput of less than $\sim$10\% at all wavelengths. Most of the errors, we believe, are due to the low S/N at the far-red end of the spectrum.
We obtained similar observations at McDonald observatory on Oct 18, 2011. McDonald is much drier compared to College Station, which makes these measurements much more comparable to CTIO, and we obtained reasonable signal through the 940nm filter. Therefore, we were able to derive the atmospheric models from both the spectroscopic measurements and the narrowband imaging. Figure \ref{fig:9} shows the best-fit model from these two different measurements. The blue dashed line is the best-fit model from the spectroscopic measurements, after comparison to all the atmospheric models in our database; the black line is selected using the measurements with the narrow-band imager, after comparison to all the synthetic color indices of each atmospheric model in our database. Two models from independent measurements shows good agreement with residuals less than 5\%.  The main errors, we think, are from the coarse grid spacing in our modeling database, which is 2mm for water and 0.05 for aerosol optical depth.  We therefore believe that the atmospheric model derived from color measurements can be an accurate representation of the true atmospheric transmission with an adequate database and a better quality imaging system.

\section{Future Work and Conclusion}
We plan to execute additional observations at CTIO in September-October 2012 with an updated prototype of aTmCam. The goal of this observing run is to finalize the observing strategy for aTmCam as it would be used as an auxiliary system for the Dark Energy Survey.  Specifically, we will measure the temporal and angular variability scales of the atmospheric transmission at CTIO over many nights in a variety of conditions. For example, if the angular variability is low or sufficiently smooth, then an eventual system would not need to co-point with the telescope for Dark Energy Survey (i.e. 4-meter Blanco Telescope) but could instead cycle around the sky to a fixed list of bright stars; that would presumably allow a smaller aperture and would decouple the system from the survey telescope.

Understanding atmospheric variation is an important step toward 1\% photometry. We proposed a relatively simple narrowband imaging system that should allow derivation of an atmospheric transmission model that could improve photometric precision to less than 1\%. We have tested a prototype of the system and confirmed, using simultaneous spectroscopic measurements, that the principle works adequately well for Vega. We plan to test our updated prototype of aTmCam at CTIO in Fall 2012, when we will also determine the temporal \& angular variality of the site.

\acknowledgements Texas A\&M University thanks Charles R. '62 and Judith G. Munnerlyn, George P. '40 and Cynthia Woods Mitchell, and their families for support of astronomical instrumentation activities in the Department of Physics and Astronomy. The author would also like to thank the support staff at Texas A\&M Observatory and McDonald Observatory for their assistance.

\bibliography{Li_T}

    \end{document}